
\font \tbfontt                = cmbx10 scaled\magstep1
\font \tafontt                = cmbx10 scaled\magstep2
\font \tbfontss               = cmbx5  scaled\magstep1
\font \tafontss               = cmbx5  scaled\magstep2
\font \sixbf                  = cmbx6
\font \tbfonts                = cmbx7  scaled\magstep1
\font \tafonts                = cmbx7  scaled\magstep2
\font \ninebf                 = cmbx9
\font \tasys                  = cmex10 scaled\magstep1

\font \sixi                   = cmmi6
\font \ninei                  = cmmi9
\font \tams                   = cmmib10
\font \tbmss                  = cmmib10 scaled 600
\font \tamss                  = cmmib10 scaled 700
\font \tbms                   = cmmib10 scaled 833
\font \tbmt                   = cmmib10 scaled\magstep1
\font \tamt                   = cmmib10 scaled\magstep2
\font \smallescriptscriptfont = cmr5
\font \smalletextfont         = cmr5 at 10pt
\font \smallescriptfont       = cmr5 at 7pt
\font \sixrm                  = cmr6
\font \ninerm                 = cmr9
\font \ninesl                 = cmsl9
\font \tensans                = cmss10
\font \fivesans               = cmss10 at 5pt
\font \sixsans                = cmss10 at 6pt
\font \sevensans              = cmss10 at 7pt
\font \ninesans               = cmss10 at 9pt
\font \tbst                   = cmsy10 scaled\magstep1
\font \tast                   = cmsy10 scaled\magstep2
\font \tbsss                  = cmsy5  scaled\magstep1
\font \tasss                  = cmsy5  scaled\magstep2
\font \sixsy                  = cmsy6
\font \tbss                   = cmsy7  scaled\magstep1
\font \tass                   = cmsy7  scaled\magstep2
\font \ninesy                 = cmsy9
\font \markfont               = cmti10 at 11pt
\font \nineit                 = cmti9
\font \ninett                 = cmtt9
\magnification=\magstep0
\hsize=13truecm
\vsize=19.8truecm
\hfuzz=2pt
\tolerance=500
\abovedisplayskip=3 mm plus6pt minus 4pt
\belowdisplayskip=3 mm plus6pt minus 4pt
\abovedisplayshortskip=0mm plus6pt minus 2pt
\belowdisplayshortskip=2 mm plus4pt minus 4pt
\predisplaypenalty=0
\clubpenalty=10000
\widowpenalty=10000
\frenchspacing
\newdimen\oldparindent\oldparindent=1.5em
\parindent=1.5em



\def\bbbc{{\mathchoice {\setbox0=\hbox{$\displaystyle\rm C$}\hbox{\hbox
to0pt{\kern0.4\wd0\vrule height0.9\ht0\hss}\box0}}
{\setbox0=\hbox{$\textstyle\rm C$}\hbox{\hbox
to0pt{\kern0.4\wd0\vrule height0.9\ht0\hss}\box0}}
{\setbox0=\hbox{$\scriptstyle\rm C$}\hbox{\hbox
to0pt{\kern0.4\wd0\vrule height0.9\ht0\hss}\box0}}
{\setbox0=\hbox{$\scriptscriptstyle\rm C$}\hbox{\hbox
to0pt{\kern0.4\wd0\vrule height0.9\ht0\hss}\box0}}}}
\def\bbbe{{\mathchoice {\setbox0=\hbox{\smalletextfont e}\hbox{\raise
0.1\ht0\hbox to0pt{\kern0.4\wd0\vrule width0.3pt
height0.7\ht0\hss}\box0}}
{\setbox0=\hbox{\smalletextfont e}\hbox{\raise 0.1\ht0\hbox
to0pt{\kern0.4\wd0\vrule width0.3pt height0.7\ht0\hss}\box0}}
{\setbox0=\hbox{\smallescriptfont e}\hbox{\raise 0.1\ht0\hbox
to0pt{\kern0.5\wd0\vrule width0.2pt height0.7\ht0\hss}\box0}}
{\setbox0=\hbox{\smallescriptscriptfont e}\hbox{\raise
0.1\ht0\hbox to0pt{\kern0.4\wd0\vrule width0.2pt
height0.7\ht0\hss}\box0}}}}
\def\bbbq{{\mathchoice {\setbox0=\hbox{$\displaystyle\rm Q$}\hbox{\raise
0.15\ht0\hbox to0pt{\kern0.4\wd0\vrule height0.8\ht0\hss}\box0}}
{\setbox0=\hbox{$\textstyle\rm Q$}\hbox{\raise
0.15\ht0\hbox to0pt{\kern0.4\wd0\vrule height0.8\ht0\hss}\box0}}
{\setbox0=\hbox{$\scriptstyle\rm Q$}\hbox{\raise
0.15\ht0\hbox to0pt{\kern0.4\wd0\vrule height0.7\ht0\hss}\box0}}
{\setbox0=\hbox{$\scriptscriptstyle\rm Q$}\hbox{\raise
0.15\ht0\hbox to0pt{\kern0.4\wd0\vrule height0.7\ht0\hss}\box0}}}}
\def\bbbt{{\mathchoice {\setbox0=\hbox{$\displaystyle\rm
T$}\hbox{\hbox to0pt{\kern0.3\wd0\vrule height0.9\ht0\hss}\box0}}
{\setbox0=\hbox{$\textstyle\rm T$}\hbox{\hbox
to0pt{\kern0.3\wd0\vrule height0.9\ht0\hss}\box0}}
{\setbox0=\hbox{$\scriptstyle\rm T$}\hbox{\hbox
to0pt{\kern0.3\wd0\vrule height0.9\ht0\hss}\box0}}
{\setbox0=\hbox{$\scriptscriptstyle\rm T$}\hbox{\hbox
to0pt{\kern0.3\wd0\vrule height0.9\ht0\hss}\box0}}}}
\def\bbbs{{\mathchoice
{\setbox0=\hbox{$\displaystyle     \rm S$}\hbox{\raise0.5\ht0\hbox
to0pt{\kern0.35\wd0\vrule height0.45\ht0\hss}\hbox
to0pt{\kern0.55\wd0\vrule height0.5\ht0\hss}\box0}}
{\setbox0=\hbox{$\textstyle        \rm S$}\hbox{\raise0.5\ht0\hbox
to0pt{\kern0.35\wd0\vrule height0.45\ht0\hss}\hbox
to0pt{\kern0.55\wd0\vrule height0.5\ht0\hss}\box0}}
{\setbox0=\hbox{$\scriptstyle      \rm S$}\hbox{\raise0.5\ht0\hbox
to0pt{\kern0.35\wd0\vrule height0.45\ht0\hss}\raise0.05\ht0\hbox
to0pt{\kern0.5\wd0\vrule height0.45\ht0\hss}\box0}}
{\setbox0=\hbox{$\scriptscriptstyle\rm S$}\hbox{\raise0.5\ht0\hbox
to0pt{\kern0.4\wd0\vrule height0.45\ht0\hss}\raise0.05\ht0\hbox
to0pt{\kern0.55\wd0\vrule height0.45\ht0\hss}\box0}}}}
\def\bbbz{{\mathchoice {\hbox{$\sans\textstyle Z\kern-0.4em Z$}}
{\hbox{$\sans\textstyle Z\kern-0.4em Z$}}
{\hbox{$\sans\scriptstyle Z\kern-0.3em Z$}}
{\hbox{$\sans\scriptscriptstyle Z\kern-0.2em Z$}}}}
\skewchar\ninei='177 \skewchar\sixi='177
\skewchar\ninesy='60 \skewchar\sixsy='60
\hyphenchar\ninett=-1
\def\newline{\hfil\break}%
\catcode`@=11
\def\folio{\ifnum\pageno<\z@
\uppercase\expandafter{\romannumeral-\pageno}%
\else\number\pageno \fi}
\catcode`@=12 
  \mathchardef\Gamma="0100
  \mathchardef\Delta="0101
  \mathchardef\Theta="0102
  \mathchardef\Lambda="0103
  \mathchardef\Xi="0104
  \mathchardef\Pi="0105
  \mathchardef\Sigma="0106
  \mathchardef\Upsilon="0107
  \mathchardef\Phi="0108
  \mathchardef\Psi="0109
  \mathchardef\Omega="010A
\def\squareforqed{\hbox{\rlap{$\sqcap$}$\sqcup$}}
\def\qed{\ifmmode\squareforqed\else{\unskip\nobreak\hfil
\penalty50\hskip1em\null\nobreak\hfil\squareforqed
\parfillskip=0pt\finalhyphendemerits=0\endgraf}\fi}
\newfam\sansfam
\textfont\sansfam=\tensans\scriptfont\sansfam=\sevensans
\scriptscriptfont\sansfam=\fivesans
\def\sans{\fam\sansfam\tensans}
\def\stackfigbox{\if
Y\FIG\global\setbox\figbox=\vbox{\unvbox\figbox\box1}%
\else\global\setbox\figbox=\vbox{\box1}\global\let\FIG=Y\fi}
\def\placefigure{\dimen0=\ht1\advance\dimen0by\dp1
\advance\dimen0by5\baselineskip
\advance\dimen0by0.4true cm
\ifdim\dimen0>\vsize\pageinsert\box1\vfill\endinsert
\else
\if Y\FIG\stackfigbox\else
\dimen0=\pagetotal\ifdim\dimen0<\pagegoal
\advance\dimen0by\ht1\advance\dimen0by\dp1\advance\dimen0by1.7true cm
\ifdim\dimen0>\pagegoal\stackfigbox
\else\box1\vskip7true mm\fi
\else\box1\vskip7true mm\fi\fi\fi\let\firstleg=Y}
%
\def\begfig#1cm#2\endfig{\par
\setbox1=\vbox{\dimen0=#1true cm\advance\dimen0
by1true cm\kern\dimen0\vskip-.8333\baselineskip#2}\placefigure}
\def\begdoublefig#1cm #2 #3 \enddoublefig{\begfig#1cm%
\line{\vtop{\hsize=0.46\hsize#2}\hfill
\vtop{\hsize=0.46\hsize#3}}\endfig}
\let\firstleg=Y
\def\figure#1#2{\if Y\firstleg\vskip1true cm\else\vskip1.7true mm\fi
\let\firstleg=N\setbox0=\vbox{\noindent\petit{\bf
Fig.\ts#1\unskip.\ }\ignorespaces #2\smallskip
\count255=0\global\advance\count255by\prevgraf}%
\ifnum\count255>1\box0\else
\centerline{\petit{\bf Fig.\ts#1\unskip.\
}\ignorespaces#2}\smallskip\fi}

\def\begtab#1cm#2\endtab{\par
   \ifvoid\topins\midinsert\medskip\vbox{#2\kern#1true cm}\endinsert
   \else\topinsert\vbox{#2\kern#1true cm}\endinsert\fi}
\def\begpet{\vskip6pt\bgroup\petit}
\def\endpet{\vskip6pt\egroup}
\newdimen\refindent
\newlinechar=`\^
\def\begref#1#2{\titlea{}{#1}%
\bgroup\petit
\setbox0=\hbox{#2\enspace}\refindent=\wd0\relax
\if>#2>\else
\ifdim\refindent>0.5em\else
\message{^Something may be wrong with your references;}%
\message{probably you missed the second argument of \string\begref.}%
\fi\fi}
\def\ref{\goodbreak
\hangindent\oldparindent\hangafter=1
\noindent\ignorespaces}
\def\refno#1{\goodbreak
\setbox0=\hbox{#1\enspace}\ifdim\refindent<\wd0\relax
\message{^Your reference `#1' is wider than you pretended in using
\string\begref.}\fi
\hangindent\refindent\hangafter=1
\noindent\kern\refindent\llap{#1\enspace}\ignorespaces}
\def\refmark#1{\goodbreak
\setbox0=\hbox{#1\enspace}\ifdim\refindent<\wd0\relax
\message{^Your reference `#1' is wider than you pretended in using
\string\begref.}\fi
\hangindent\refindent\hangafter=1
\noindent\hbox to\refindent{#1\hss}\ignorespaces}
\def\endref{\goodbreak\endpet}
\def\petit{\def\rm{\fam0\ninerm}%
\textfont0=\ninerm \scriptfont0=\sixrm \scriptscriptfont0=\fiverm
 \textfont1=\ninei \scriptfont1=\sixi \scriptscriptfont1=\fivei
 \textfont2=\ninesy \scriptfont2=\sixsy \scriptscriptfont2=\fivesy
 \def\it{\fam\itfam\nineit}%
 \textfont\itfam=\nineit
 \def\sl{\fam\slfam\ninesl}%
 \textfont\slfam=\ninesl
 \def\bf{\fam\bffam\ninebf}%
 \textfont\bffam=\ninebf \scriptfont\bffam=\sixbf
 \scriptscriptfont\bffam=\fivebf
 \def\sans{\fam\sansfam\ninesans}%
 \textfont\sansfam=\ninesans \scriptfont\sansfam=\sixsans
 \scriptscriptfont\sansfam=\fivesans
 \def\tt{\fam\ttfam\ninett}%
 \textfont\ttfam=\ninett
 \normalbaselineskip=11pt
 \setbox\strutbox=\hbox{\vrule height7pt depth2pt width0pt}%
 \normalbaselines\rm
\def\vec##1{{\textfont1=\tbms\scriptfont1=\tbmss
\textfont0=\ninebf\scriptfont0=\sixbf
\mathchoice{\hbox{$\displaystyle##1$}}{\hbox{$\textstyle##1$}}
{\hbox{$\scriptstyle##1$}}{\hbox{$\scriptscriptstyle##1$}}}}}
\nopagenumbers
%
\let\header=Y
\let\FIG=N
\newbox\figbox
\output={\if N\header\headline={\hfil}\fi\plainoutput
\global\let\header=Y\if Y\FIG\topinsert\unvbox\figbox\endinsert
\global\let\FIG=N\fi}
\let\lasttitle=N
\def\centerpar#1{{\parfillskip=0pt
\rightskip=0pt plus 1fil
\leftskip=0pt plus 1fil
\advance\leftskip by\oldparindent
\advance\rightskip by\oldparindent
\def\newline{\break}%
\noindent\ignorespaces#1\par}}
\catcode`\@=\active
\def\author#1{\bgroup
\baselineskip=13.2pt
\lineskip=0pt
\pretolerance=10000
\markfont
\centerpar{#1}\bigskip\egroup
{\def@##1{}%
\setbox0=\hbox{\petit\kern2.5true cc\ignorespaces#1\unskip}%
\ifdim\wd0>\hsize
\message{The names of the authors exceed the headline, please use a }%
\message{short form with AUTHORRUNNING}\gdef\leftheadline{%
\rlap{\folio}\hfil AUTHORS suppressed due to excessive length}%
\else
\xdef\leftheadline{\rlap{\noexpand\folio}\hfil
\ignorespaces#1\unskip}%
\fi
}\let\INS=E}
\def\address#1{\bgroup\petit
\centerpar{#1}\bigskip\egroup
\catcode`\@=12
\vskip2cm\noindent\ignorespaces}
\let\INS=N%
\def@#1{\if N\INS\unskip$\,^{#1}$\else\global\footcount=#1\relax
\if E\INS\hangindent0.5\parindent\noindent\hbox
to0.5\parindent{$^{#1}$\hfil}\let\INS=Y\ignorespaces
\else\par\hangindent0.5\parindent\noindent\hbox
to0.5\parindent{$^{#1}$\hfil}\ignorespaces\fi\fi}%
\catcode`\@=12
\headline={\petit\def\newline{ }\def\fonote#1{}\ifodd\pageno
\rightheadline\else\leftheadline\fi}
\def\rightheadline{Missing CONTRIBUTION
title\hfil\llap{\folio}}
\def\leftheadline{\rlap{\folio}\hfil Missing name(s)
of the author(s)}
\nopagenumbers
\let\header=Y

\let\lasttitle=N
 \def\contribution#1{\vfill\eject
 \let\header=N\bgroup
 \textfont0=\tafontt \scriptfont0=\tafonts \scriptscriptfont0=\tafontss
 \textfont1=\tamt \scriptfont1=\tams \scriptscriptfont1=\tams
 \textfont2=\tast \scriptfont2=\tass \scriptscriptfont2=\tasss
 \par\baselineskip=16pt
     \lineskip=16pt
     \tafontt
     \raggedright
     \pretolerance=10000
     \noindent
     \centerpar{\ignorespaces#1}%
     \vskip17pt\egroup
     \nobreak
     \parindent=0pt
     \everypar={\global\parindent=1.5em
     \global\let\lasttitle=N\global\everypar={}}%
     \global\let\lasttitle=A%
     \setbox0=\hbox{\petit\def\newline{ }\def\fonote##1{}\kern2.5true
     cc\ignorespaces#1}\ifdim\wd0>\hsize
     \message{Your CONTRIBUTIONtitle exceeds the headline,
please use a short form
with CONTRIBUTIONRUNNING}\gdef\rightheadline{CONTRIBUTION title
suppressed due to excessive length\hfil\llap{\folio}}%
\else
\gdef\rightheadline{\ignorespaces#1\unskip\hfil\llap{\folio}}\fi
\catcode`\@=\active
     \ignorespaces}
\def\titlea#1#2{\if N\lasttitle\else\vskip-28pt
     \fi
     \vskip18pt plus 4pt minus4pt
     \bgroup
\textfont0=\tbfontt \scriptfont0=\tbfonts \scriptscriptfont0=\tbfontss
\textfont1=\tbmt \scriptfont1=\tbms \scriptscriptfont1=\tbmss
\textfont2=\tbst \scriptfont2=\tbss \scriptscriptfont2=\tbsss
\textfont3=\tasys \scriptfont3=\tenex \scriptscriptfont3=\tenex
     \baselineskip=16pt
     \lineskip=0pt
     \pretolerance=10000
     \noindent
     \tbfontt
     \rightskip 0pt plus 6em
     \setbox0=\vbox{\vskip23pt\def\fonote##1{}%
     \noindent
     \if>#1>\ignorespaces#2
     \else\ignorespaces#1\unskip\enspace\ignorespaces#2\fi
     \vskip18pt}%
     \dimen0=\pagetotal\advance\dimen0 by-\pageshrink
     \ifdim\dimen0<\pagegoal
     \dimen0=\ht0\advance\dimen0 by\dp0\advance\dimen0 by
     3\normalbaselineskip
     \advance\dimen0 by\pagetotal
     \ifdim\dimen0>\pagegoal\eject\fi\fi
     \noindent
     \if>#1>\ignorespaces#2
     \else\ignorespaces#1\unskip\enspace\ignorespaces#2\fi
     \vskip12pt plus4pt minus4pt\egroup
     \nobreak
     \parindent=0pt
     \everypar={\global\parindent=\oldparindent
     \global\let\lasttitle=N\global\everypar={}}%
     \global\let\lasttitle=A%
     \ignorespaces}
 \def\titleb#1#2{\if N\lasttitle\else\vskip-22pt
     \fi
     \vskip18pt plus 4pt minus4pt
     \bgroup
\textfont0=\tenbf \scriptfont0=\sevenbf \scriptscriptfont0=\fivebf
\textfont1=\tams \scriptfont1=\tamss \scriptscriptfont1=\tbmss
     \lineskip=0pt
     \pretolerance=10000
     \noindent
     \bf
     \rightskip 0pt plus 6em
     \setbox0=\vbox{\vskip23pt\def\fonote##1{}%
     \noindent
     \if>#1>\ignorespaces#2
     \else\ignorespaces#1\unskip\enspace\ignorespaces#2\fi
     \vskip10pt}%
     \dimen0=\pagetotal\advance\dimen0 by-\pageshrink
     \ifdim\dimen0<\pagegoal
     \dimen0=\ht0\advance\dimen0 by\dp0\advance\dimen0 by
     3\normalbaselineskip
     \advance\dimen0 by\pagetotal
     \ifdim\dimen0>\pagegoal\eject\fi\fi
     \noindent
     \if>#1>\ignorespaces#2
     \else\ignorespaces#1\unskip\enspace\ignorespaces#2\fi
     \vskip8pt plus4pt minus4pt\egroup
     \nobreak
     \parindent=0pt
     \everypar={\global\parindent=\oldparindent
     \global\let\lasttitle=N\global\everypar={}}%
     \global\let\lasttitle=B%
     \ignorespaces}
 \def\titlec#1{\if N\lasttitle\else\vskip-\baselineskip
     \fi
     \vskip18pt plus 4pt minus4pt
     \bgroup
\textfont0=\tenbf \scriptfont0=\sevenbf \scriptscriptfont0=\fivebf
\textfont1=\tams \scriptfont1=\tamss \scriptscriptfont1=\tbmss
     \bf
     \noindent
     \ignorespaces#1\unskip\ \egroup
     \ignorespaces}
 \def\titled#1{\if N\lasttitle\else\vskip-\baselineskip
     \fi
     \vskip12pt plus 4pt minus 4pt
     \bgroup
     \it
     \noindent
     \ignorespaces#1\unskip\ \egroup
     \ignorespaces}
\let\ts=\thinspace
\def\footnoterule{\kern-3pt\hrule width 2true cm\kern2.6pt}
\newcount\footcount \footcount=0
\def\advftncnt{\advance\footcount by1\global\footcount=\footcount}
\def\fonote#1{\advftncnt$^{\the\footcount}$\begingroup\petit
\parfillskip=0pt plus 1fil
\def\textindent##1{\hangindent0.5\oldparindent\noindent\hbox
to0.5\oldparindent{##1\hss}\ignorespaces}%
\vfootnote{$^{\the\footcount}$}{#1\vskip-9.69pt}\endgroup}
\def\item#1{\par\noindent
\hangindent6.5 mm\hangafter=0
\llap{#1\enspace}\ignorespaces}

\def\newenvironment#1#2#3#4{\long\def#1##1##2{\removelastskip
\vskip\baselineskip\noindent{#3#2\if>##1>.\else\unskip\ \ignorespaces
##1\unskip\fi\ }{#4\ignorespaces##2}\vskip\baselineskip}}
\newenvironment\lemma{Lemma}{\bf}{\it}
\newenvironment\proposition{Proposition}{\bf}{\it}
\newenvironment\theorem{Theorem}{\bf}{\it}
\newenvironment\corollary{Corollary}{\bf}{\it}
\newenvironment\example{Example}{\it}{\rm}
\newenvironment\exercise{Exercise}{\bf}{\rm}
\newenvironment\problem{Problem}{\bf}{\rm}
\newenvironment\solution{Solution}{\bf}{\rm}
\newenvironment\definition{Definition}{\bf}{\rm}
\newenvironment\note{Note}{\it}{\rm}
\newenvironment\question{Question}{\it}{\rm}
\long\def\remark#1{\removelastskip\vskip\baselineskip\noindent{\it
Remark.\ }\ignorespaces}
\long\def\proof#1{\removelastskip\vskip\baselineskip\noindent{\it
Proof\if>#1>\else\ \ignorespaces#1\fi.\ }\ignorespaces}
\def\typeset{\petit\noindent This article was processed by the author
using the \TeX\ macro package from Springer-Verlag.\par}
\outer\def\byebye{\bigskip\bigskip\typeset
\footcount=1\ifx\speciali\undefined\else
\loop\smallskip\noindent special character No\number\footcount:
\csname special\romannumeral\footcount\endcsname
\advance\footcount by 1\global\footcount=\footcount
\ifnum\footcount<11\repeat\fi
\vfill\supereject\end}

\def\12{{1\ov 2}}

\def\ov{\over}


\contribution{Threshold Photo/Electro Pion Production -- Working Group Summary}
\author{Ulf-G. Mei{\ss}ner@1, B. Schoch@2}\footnote{}{\hskip -0.6truecm
Summary talk presented at the Workshop on Chiral Dynamics: Theory and
Experiments, Massachusetts Institute of Technology, Cambridge,
USA, July 25 - 29, 1994, CRN 94-51}
\address{@1Centre de Recherches Nucl\'eaire, Physique Th\'eorique, BP 28 Cr,
F--67037 Strasbourg Cedex 2, France @2Universit\"at Bonn, Institut f\"ur
Physik, Nu{\ss}allee 12, D--53115 Bonn, Germany}
\vskip -2truecm
\titlea{1}{Introduction}
Over the last few years, pion production off nucleons by real or virtual
photons has become a central issue in the study of the non--perturbative
structure of the nucleon, i.e. at low energies. Here, developments in detector
and accelerator technology on the experimental side as well as better
calculational tools on the theoretical one have allowed to gain more insight
into detailed aspects of these processes and the physics behind them. One main
trigger were the two papers by the Saclay and the Mainz groups [1,2],
which seemed to indicate the violation of a so--called low energy theorem for
the reaction $\gamma p \to \pi^0 p$.  This lead to a flurry of further
experimental and theoretical investigations. Another cornerstone was the rather
precise electroproduction measurement $\gamma^\star p \to \pi^0 p$ at NIKHEF
[3]. Here, we wish to summarize the state of the art in calculating and
measuring these processes in the threshold region. Furthermore, we outline what
we believe have crystalized as the pertinent activities to be done in the near
future.

\titlea{2}{Theoretical developments}
The chiral perturbation theory (CHPT) machinery to calculate the reactions
$\gamma N \to \pi^a N$, $\gamma^\star N \to \pi^a N$ and
$\gamma^{(\star)} N \to \pi^a \pi^b N$, where $N$ denotes the nucleon, $\pi^a$
a pion of isospin $'a'$ and $\gamma \, (\gamma^\star )$ the real (virtual)
photon exists as described in some detail in V. Bernard's lecture [4]. It has
become clear that to have precise predictions one has to calculate
in the one loop approximation to order
$q^4$ in the effective Lagrangian,
$${\cal L}_{\rm eff} =  {\cal L}_{\pi N}^{(1)} + {\cal L}_{\pi N}^{(2)}
+ {\cal L}_{\pi N}^{(3)} + {\cal L}_{\pi N}^{(4)}
\eqno(2.1)$$
where the subsript $' \pi N '$ means that we restrict ourselves to the two
flavor case (the pion--nucleon--photon system) and the superscript $'( i) ' $
denotes the chiral dimension.  While the first term in eq.(2.1) is given
entirely in terms of well determined parameters, the string of terms of order
$q^2, q^3$ and $q^4$ contains the so--called low--energy constants (LECs).
At present,
their determination induces the biggest uncertainty in the chiral predictions
since not enough sufficiently accurate low--energy data exist to uniquely pin
them down. Therefore, in the case of threshold photo/electro pion production,
we follow two approaches. The first one is "clean" CHPT, were the extensive
photoproduction data (total and differential cross
sections in the threshold region) are used to
determine the appearing LECs (see the discussion in Bernard's lecture [4]). In
that case, the predictive power resides in the photoproduction multipoles,
polarization observables and the electroproduction processes. For the latter,
some novel LECs appear, but they are all connected to single nucleon
properties and thus can uniquely be fixed from the electromagnetic and axial
radii of the nucleon. With that, the $k^2$ dependence (where $k^2 <0$ denotes
the photon four--momentum squared) of all electroproduction observables and
multipoles is {\it parameter--free} fixed.
 For detailed calculations and predictions to order
$q^3$ consult the {\it Physics Reports} [5].
Furthermore, one gains new insight due
to the longitudinal coupling of the virtual photon to the nucleon spin. This
rich field has only be glimpsed at and will serve as a good testing ground
for the chiral predictions.

Now let us discuss what the limitations of such calculations are, i.e. to what
values of $|k^2|$ and $\Delta W = W - W_{\rm thr}$ (with $W$ the
center--of--mass energy of the $\pi N$ system) these calcualtions to order
$q^4$ can be trusted? The answer is, of course, dependent on the observable one
looks at. Nevertheless, a good example is given by the one loop calculation of
the S--wave cross section $a_0$ in $\gamma^\star p \to \pi^0 p$ [6] in
comparison to the data of ref.[3]. Although the trend of the data is well
 reproduced up to $k^2 = -0.1$ GeV$^2$, it is obvious that for such large
four--momentum transfer ($|k| \simeq 2.3 M_\pi$) the one--loop corrections to
the tree result are so large that they can not be trusted quantitatively, i.e.
one would have to calculate further in the chiral expansion. From the extensive
study of elastic pion--pion scattering in the threshold region[7], which is the
purest process to test the chiral dynamics, we employ here the same rule of
thumb advocated there, namely that as long as the one--loop corrections stay
below 50$\%$ of the tree result, the predictions can be considered quantitative
(accurate). For the S--wave cross section $a_0$ and many other observables, we
conclude that for a rigorous test of the chiral dynamics, electroproduction
experiments should be performed with
$$ |k^2| \leq 0.05 \, {\rm GeV}^2 \, \quad \Delta W \leq 15 \, {\rm MeV}
\eqno(2.2)$$
where the number for $\Delta W$ is derived from photoproduction calculations
and deserves more detailed studies. In any case, it is conceivable that one
should stay as close to threshold (and the photon point) as experimentally
possible.

The second approach to get a handle on the LECs involves the principle of
resonance saturation (as discussed in some detail
in these proceedings by Ecker [8] and by  one of
us [9]). Consider e.g. a contact term of order $q^3$ with an incoming photon
and
one outgoing pion. The resonance saturation hypothesis means that the numerical
value of the LEC related to this operator is given by baryon excitations like
the $\Delta (1232)$, the $N^\star (1440)$ and others (as discussed here e.g.
by Mukhopadhyay [10]) and also t--channel meson exchanges of scalar, vector or
axial--vector type. Shrinking such types of diagrams to a point, the pertinent
LEC is given in terms of masses and coupling constants related to the particles
integrated out from the effective field theory. This, of course, induces some
model--dependence but can on the other hand serve as a guideline to understand
the numerical values of the LECs from some microscopic picture. However, we
would like to stress again that it is preferable to have {\it enough} data to
pin down all these low--energy constants, thereby testing the accurary of the
resonance saturation idea (which works very well for strong and semi--leptonic
interactions in the meson sector, but not for the non--leptonic weak
interactions).  This principle could then be used to
estimate the contributions of higher order terms as it is frequently done in
the meson sector. Also, we wish to point out that all this resonance physics
seen in the threshold region is, of course, included in the phenomenological
LECs. As long as one is not so ambitious to make statements throughout or above
the region of nucleon excitations, {\it resonance contributions do not pose
any problem to the consistent chiral expansion}.

We would also like to stress the importance of polarization observables. These
can be calculated with ease and they serve as a sensitive filter in the
multipole analysis which otherwise is difficult to perform and often not unique
(in the sense that one has to determine more multipoles than the number of
available observables). Another important topic is the neutron. Although no
free neutron targets are at our disposal, measurements using the deuteron can
serve as neutron probes. For that, it is mandatory to have a precise
description of the weakly bound proton--neutron system. This can e.g. be
achieved in the relativistic model of Gross and collaborators [11]. Clearly,
some model--dependence is induced, but one hopes to be able to minimize this by
a) measuring also the proton in the deuteron and comparing to its free space
values and b) chosing particular kinematics where the corrections due to meson
exchange currents and alike are minimized.

What are the theoretical improvements needed? First, in certain channels (where
one has strong final--state interactions) it is mandatory to perform
calculations including higher order effects. The most efficient machinery to do
this is a combination of dispersion theory with CHPT as discussed in detail for
pion form factors or $K_{\ell 4}$--decays in refs.[12]. In essence, consider an
observable ${\cal Q}$ and write a dispersive representation,
$$ {\cal Q} = P (s) \, \Omega_\Lambda (s)            \eqno(2.3)$$
where the reduced Omn{\`e}s function $ \Omega_\Lambda (s) $ accounts for the
phase information (and sums up loops to very high orders), the polynomial
$P(s)$ is constrained by chiral symmetry and $s$ is a short--hand notation for
the kinematical variables involved. Such a method has e.g. been applied
succesfully in the determination of the $\pi N$ $\sigma$--term and the scalar
form factor of the nucleon as discussed by Sainio [13]. Also, it is an open
question to what extent one should include the $\Delta (1232)$ as a dynamical
degree of freedom in the effective field theory.  We believe that for the
threshold phenomena to be mapped out by the experiments within the constraints
of eq.(2.2), it suffices to use the $\Delta$ to estimate LECs and thus treat it
as a frozen (integrated out) d.o.f.  The most important theoretical issue to be
studied in more detail is the role of isospin breaking through virtual photons
and the quark mass difference $m_u - m_d$. At present, one puts in the
different masses for the neutral and charged pions by hand in a manner
consistent with all the
symmetries and Ward identities. It is well--known that the pion mass differnec
is essentially of electromagnetic origin, leading to the believe that this
approximation accounts for the most important aspects of isospin breaking.
Nevertheless, this has to be clarified in a  detailed and thorough study. In
any case, threshold pion photo-- and electroproduction can not be considered
alone but it is intimately connected to other processes like Compton
scattering, pion--nucleon scattering, $\pi N \to \pi \pi N$ and so on. This
interplay of the various reactions is of particular importance for the
determinations of the LECs.

Finally, a few words about the extension to three flavors are in order. Here,
the theoretical situation is much less satisfactory mostly due to the large
kaon loop contributions and huge cancellations with counterterms. Stated
differently, the intrinsic small parameter is $(M_K/4\pi F_\pi)^2 \simeq
(0.4)^2$ (modulo logarithms) and
thus it is mandatory to perform higher order calculations than for the two
flavor case. This becomes particularly evident in the scalar sector of the
baryon masses and $\sigma$--terms as detailed in ref.[14]. As long as a
consistent theoretical description of these 2-- and 3--point functions is
eluding us, we can not make statements about kaon and eta production off the
nucleon. However, beautiful new data for $K$ and $\eta$ production in the
threshold region are becoming available ($\gamma p \to \eta p$, $\gamma p \to
\Sigma K$ or $\gamma p \to \Lambda K$) which should lead to more theoretical
effort to understand the chiral SU(3) meson--baryon system.

\titlea{3}{Experimental developments}

Driven by the earlier results [1], [2] of the
$\pi^0$--production at threshold on the proton and the discussion about
the validity of low energy theorems experimental activities
started in several laboratories in order to study the
photo--and electroproduction of charged and neutral pions. The
experiment of the electroproduction of neutral pions on the
proton at NIKHEF [3] demonstrated the possibility to perform
these experiments with the precision to be able to discriminate
between different theoretical approaches. H. Blok, Vrije
Universiteit Amsterdam, showed preliminary results of the $^1H
(e,e'p) \pi^0$--reaction at a momentum transfer of
$|k|^2=0.1$ (GeV/c)$^2$, thereby, covering an invariant mass range of
2 to 15 MeV above the production threshold. M. Distler,
Universit\"at Mainz, presented the first data of the $^1H (e,e'p)
\pi^0$--reaction with the new spectrometer set-up at MAMI.
Measurements at two polarizations of the virtual photon
$(\varepsilon_1 = 0.89$ and $\varepsilon_2 = 0.52)$ allow a separation
of the transverse and longitudinal contributions.

Due to the large acceptance in the threshold region the
out--of--plane cross sections $\sigma_{TL}$ and $\sigma_{TT}$ can be
determined. The momentum transfer of $|k|^2 =0.1$ (GeV/C)$^2$ like in
the NIKHEF--experiment exceeds, however, (see equation 2.2) the
region of the validity of present calculations within the
framework of CHPT. Nevertheless, for the first time these high
quality data will allow the extraction of the S-- and P--wave
multipoles. In addition to these measurements results of the
electroproduction of positive pions from an experiment carried
out at MAMI have been announced. Besides the possibility to
break down the amplitudes into their isospin pieces the
extraction of the axial formfactor at low $|k|^2$ becomes
possible. Again, the results of calculations based on CHPT can
be checked and compared with results of neutrino induced
reactions.

J. Bergstroem, Saskatoon, reported about an ongoing experiment
$^1H ( \gamma , \pi^0)$ at SAL (Saskatoon) and R. Beck (Mainz)
described a planned experiment of the reaction $^1H
(\vec{\gamma} , \pi^0)$ at threshold. This experiment will allow
a clean separation of the S-- and P--wave contribution to the
cross section in the region below and above the threshold of
positive pions.

The electric amplitudes of the charged channels are fixed to the
first order by the Kroll--Rudermann term. So far, the published
experimental data agree within a few percent with this
prediction. However, in the $\pi^+$--case there exist no data of
the differential cross section close to threshold. Either an
extrapolation over more than 5 MeV was performed or the
knowledge about the P--wave was used in order to extract the
electric multipole from the measurement of the total cross
section. J. Bergstroem reported about a recent experiment of the
production of negative pions on the deuteron from SAL. F. Klein
discussed results for this reaction from Mainz and addressed the
problems which arise in order to get information for the
elementary amplitude on the neutron.

These problems do not arise by investigating the inverse
reaction $\pi^- p \rightarrow \gamma n$. M. Kovash presented
new, again preliminary data from TRIUMF which have the accuracy
to check the electric amplitude for negative pion production
extracted, so far, from a measurement of the Panofski--ratio. As
a by--product of this measurement the Panofski--ratio was
measured, simultaneously, and, given the impressive quality of
the raw data, an improved value for this ratio should be
obtained as Pavan demonstrated in reviewing the previous
measurements.

These new experimental results demonstrate the renewed interest
for the physics questions and show the impressive progress in
the experimental techniques.

On this basis even more difficult experiments can be envisaged.
The measurement of the neutral pion production on the neutron
constitutes such an experiment which would, combined with the
older results, check the validity of the isospin decomposition
of the production amplitudes. In this connection A. Bernstein
proposed a measurement using a polarized beam and a polarized
target in order to measure very close to threshold the phase
shift of the $\pi^0-p$ scattering process. Precise measurements
on the proton and the neutron would yield a high sensitivity for
isospin--breaking effects due to the difference of the current
mass of the u-- and d--quarks. High precision data from MAMI and
ELSA (Bonn) at the $\eta$--threshold are now available for the
proton as well as for the neutron. Also data for the
kaon--production have been extracted from measurements with the
SAPHIR-detector on ELSA. Besides total and differential cross
sections the measurement of polarization observables allow in
the near future a decomposition in all multipoles. The
difficulties to describe these reactions within the framework of
CHPT have been addressed in chapter 2.

\titlea{4}{Concluding remarks}

Remarkable progress has been achieved in recent years concerning
the understanding of the pion production at threshold. Due to
the progress in the development of CHPT the foundations of Low
Energy Theorems have been revisited and put into a new
perspective (for a tutorial, see e.g. ref.[15]).
CHPT provides a natural link for {\it all} low energy
pion reactions. The development of a new generation of electron
accelerators in combination with refined detector techniques
allow precision experiments, a precondition to explore the implications
of the spontaneously broken chiral symmetry of QCD in the baryon sector.

\titlea{5}{Acknowledgements}

We would like to thank everybody in the working group for presenting their
ideas and stimulating discussions.

\begref{References}{[MT1]}
\refno{[1]}E. Mazzucato et al., {\it Phys. Rev. Lett.\/} {\bf 57} (1986) 3144
\refno{[2]}R. Beck et al., {\it Phys. Rev. Lett.\/} {\bf 65} (1990) 1841
\refno{[3]}T. P. Welch et al., {\it Phys. Rev. Lett.\/} {\bf 69} (1992) 2761
\refno{[4]}V. Bernard, these proceedings
\refno{[5]}V. Bernard, N. Kaiser, T.--S. H. Lee and Ulf-G. Mei{\ss}ner,
{\it Phys. Reports\/}  (1994),  in print
\refno{[6]}
V. Bernard, N. Kaiser, T.-S.H. Lee and Ulf-G. Mei{\ss}ner,
{\it Phys. Rev. Lett.} {\bf 70} (1993) 387
\refno{[7]}J. Gasser and Ulf-G. Mei{\ss}ner,
{\it Phys. Lett.\/} {\bf B258} (1991) 219
\refno{[8]}G. Ecker, these proceedings
\refno{[9]}Ulf-G. Mei{\ss}ner, these proceedings
\refno{[10]}R.M. Davidson and N.C. Mukhopadhyay, {\it Phys. Rev. Lett.} {\bf
60} (1988) 746; M. Benmerrouche and N.C. Mukhopadhyay, {\it ibid},
{\bf 67} (1991) 1070; M. Benmerrouche, N.C. Mukhopadhyay and J.F. Zhang,
submitted to {\it Phys. Rev.} {\bf D}
\refno{[11]}F. Gross and D.O. Riska, {\it Phys. Rev.} {\bf C36} (1987) 1928;
F. Gross, {\it Czech. J. Phys.} {\bf B39} (1989) 871;
J.W. van Orden, F. Gross and N. Devine, in preparation
\refno{[12]}J. Gasser and Ulf-G. Mei{\ss}ner,
{\it Nucl. Phys.\/} {\bf B357} (1991) 91; J. Bijnens et al., BUTP--94/4,
to appear
\refno{[13]}M.E. Sainio, these proceedings
\refno{[14]} V. Bernard, N. Kaiser and Ulf-G. Mei{\ss}ner,
{\it Z. Phys.\/} {\bf C60} (1993) 111
\refno{[15]} G. Ecker and Ulf-G. Mei{\ss}ner, "What is a Low--Enegy Theorem?",
preprint CRN 94/51 and UWThPh-1994-33 (1994)
\endref
\byebye